\documentclass[journal,transmag]{IEEEtran}

\usepackage[pdftex]{graphicx}
\usepackage{amsmath}
\usepackage{bm}
\usepackage{tuda-pgfplots}
\usepackage{tablefootnote}
\usepackage{siunitx}
\usepackage[font=footnotesize]{caption}
\usepackage{subcaption}
\usepackage{amsfonts}
\usepackage{tikz}
\usepackage{pgfplots}
\usepackage{cite}
\usepackage{xcolor}
\usepackage{comment}
\usepackage{mathtools}
\usepackage{hyperref}
\usepackage{eso-pic}
\hypersetup{hidelinks}
\pgfplotsset{compat=1.18}

\graphicspath{ {./images/} }
\DeclareGraphicsExtensions{.pdf,.jpeg,.png,.tikz}

\makeatletter
\patchcmd{\@makecaption}
    {\scshape}
    {}
    {}
    {}
\makeatother

\newcommand{\mb}[1]{\textcolor{black}{#1}}
\newcommand{\rwi}[1]{\textcolor{black}{#1}}
\newcommand{\rwii}[1]{\textcolor{black}{#1}}
\newcommand{\rwiii}[1]{\textcolor{black}{#1}}

\AddToShipoutPicture*{
	\footnotesize\sffamily\raisebox{0.8cm}{\hspace{1.4cm}\fbox{
			\parbox{\textwidth}{
				© 2025 IEEE. Personal use of this material is permitted. Permission from IEEE must be obtained for all other uses, including reprinting/republishing this material for advertising or promotional purposes, collecting new collected works for resale or redistribution to servers or lists, or reuse of any copyrighted component of this work in other works.
			}
	}}
}
\begin{document}

\title{Learning electromagnetic fields based on finite element basis functions}
\author{\IEEEauthorblockN{Merle Backmeyer\IEEEauthorrefmark{1,2},
Michael Wiesheu\IEEEauthorrefmark{1}, and
Sebastian Schöps\IEEEauthorrefmark{1}}
\IEEEauthorblockA{\IEEEauthorrefmark{1}Computational Electromagnetics Group, Technische Universität Darmstadt, 64289 Darmstadt, Germany \\
\IEEEauthorrefmark{2} Univ. Grenoble Alpes, CNRS, Grenoble INP\footnote{Institute of Engineering Univ. Grenoble Alpes}, G2Elab, 38000 Grenoble, France}%
\thanks{%
    Received July 2025; revised October 2025; accepted November 2025. Date of publication November 2025; date of current version October 2025. Corresponding author: M. Backmeyer (email: merle.backmeyer@tu-darmstadt.de).%
}
\thanks{%
    Color versions of one or more figures in this article are available at
https://doi.org/10.1109/TMAG.2025.3629546.%
}
\thanks{%
    Digital Object Identifier 10.1109/TMAG.2025.3629546.%
}
}

\markboth{Learning electromagnetic fields based on finite element basis functions}%
{Learning electromagnetic fields based on finite element basis functions}

\IEEEtitleabstractindextext{%
\begin{abstract}
Parametric surrogate models of electric machines are widely used for efficient design optimization and operational monitoring. Addressing geometry variations, spline-based computer-aided design representations play a pivotal role. In this study, we propose a novel approach that combines isogeometric analysis, proper orthogonal decomposition and deep learning to enable rapid and physically consistent predictions by directly learning spline basis coefficients. The effectiveness of this method is demonstrated using a parametric nonlinear magnetostatic model of a permanent magnet synchronous machine.
\end{abstract}

\begin{IEEEkeywords}
Isogeometric Analysis, Physics-Informed Neural Networks, Permanent Magnet Synchronous Motor
\end{IEEEkeywords}}

\maketitle
\IEEEdisplaynontitleabstractindextext
\IEEEpeerreviewmaketitle

\section{Introduction}

\IEEEPARstart{D}{esign} and optimization of electric machines, such as Permanent Magnet Synchronous Motors (PMSMs), demand fast yet accurate electromagnetic field solutions for many parameter configurations. This need arises in contexts like design space exploration, uncertainty quantification, and real-time monitoring. 

In recent years, data-driven surrogate models have gained significant interest as a way to circumvent this bottleneck~\cite{Alizadeh_2020aa}. While numerous studies opt to bypass intermediate field descriptions, focusing instead on learning key performance indicators directly, e.g. \cite{Parekh_2023aa}, we advocate to retain the full field solution as the surrogate’s output. This has two main advantages. First, many Quantities of Interest (QoIs) are linear or quadratic functionals of the fields, so they are cheap to evaluate and inherit the physical rigor embedded in the field representation \cite{Salon_1995aa,Rosu_2017aa}. For example, an energy remains meaningful and positive due to its quadratic structure. Second, keeping the field solution enables the evaluation of multiple QoIs without retraining, supporting a broad range of other design and control tasks.

Physics-Informed Neural Networks (PINNs) are among the prominent approaches to learn field representations. They enforce the governing partial differential equations (PDEs) as soft constraints during training~\cite{Raissi_2019aa}. Originally developed to solve single instances, PINNs have since been extended to parametric problems, promising rapid predictions for new parameter values through a simple network evaluation~\cite{Beltran-Pulido_2022aa}. However, ensuring strict physical consistency in the resulting solutions -- particularly across material interfaces -- remains challenging for PINNs, and their training can become demanding for complex electromagnetic problems, e.g.~\cite{von-Tresckow_2022aa}.

In contrast, finite element analysis (FEA) naturally respect such conditions through carefully constructed basis functions that guarantee continuity or discontinuity where required~\cite{Monk_2003aa}.
Recently, the idea arose to combine these worlds by training neural networks to predict the coefficients of the solution with respect to the discretization basis rather than the solution fields directly. In electromagnetism, this coefficient-based learning concept has been explored for both low-order FEA and high-order isogeometric analysis (IGA) bases, with contributions by Zorzetto~\cite{Zorzetto_2025aa} for FE and with Gaussian Processes as well as Möller~\cite{Moller_2021aa} and Backmeyer~\cite{Backmeyer_2024ab} for IGA with Deep Neural Networks (DNNs).
IGA is particularly attractive since its spline representation yields more accurate approximations for the same number of degrees of freedom (DoFs) as classical FE \cite{Evans_2009aa}. As a result, to achieve a given level of accuracy, learning the coefficients with respect to an IGA basis typically requires fewer DoFs, which reduces the training complexity and data requirements. In addition, IGA provides exact representations of conic sections and enables the modeling of large deformations without remeshing due to patch-wise NURBS geometry description \cite{Cohen_2001aa,Buffa_2015aa}. This means that coefficients and basis functions can be straightforwardly learned in the reference domain.

Although IGA already reduces the output space, further compression can make learning more efficient. By projecting the solution onto a problem-specific proper orthogonal decomposition (POD) basis \cite{Benner_2017aa}, the network only needs to learn the dominant modes. This is conceptually similar to the approaches presented in \cite{Swischuk_2019aa,Henneron_2020aa, Fresca_2022aa}.
The approximation error introduced by the POD can be made negligible compared to the DNN prediction error. This reduces the output dimension the network must learn, improving generalization, lowering data requirements, and speeding up training. While nonlinear autoencoders can also provide compact latent spaces, Zorzetto et al.~\cite{Zorzetto_2025ab} found that POD offers comparable accuracy with the added benefits of interpretability, reproducibility, and straightforward implementation. In this work, we adopt a POD with an appropriately weighted inner product, chosen to be consistent with the solution space of the IGA discretization. This weighting ensures that the resulting modes reflect the relevant structure of the solution, properly account for non-uniform meshes \cite{Volkwein_2013aa}.

The proposed POD-DNN surrogate is demonstrated for a permanent magnet synchronous machine (PMSM). To highlight the flexibility of the approach, we explore two variants: first, a surrogate that learns only the air gap field distribution, which is sufficient for computing certain quantities of interest such as the torque; and second, an extended version that predicts the full magnetic field solution in rotor, stator, and air gap regions. The surrogate can reconstruct the value of the magnetic field in the PMSM for varying geometric parameters, including the rotor position, \mb{design parameters of the magnets and flux barriers, as well as driving currents,} demonstrating its potential to deliver physically consistent field predictions with minimal computational effort. 

The remainder of this paper is structured as follows: \autoref{sec:model} introduces the physical problem and its geometric parametrization, Section~\ref{sec:methodology} details the surrogate modeling framework, Section~\ref{sec:results} presents numerical results, and Section~\ref{sec:conclusion} concludes the work.

\section{Model Problem}

\label{sec:model}
\begin{figure}
	\centering
\scalebox{0.45}{
    \input{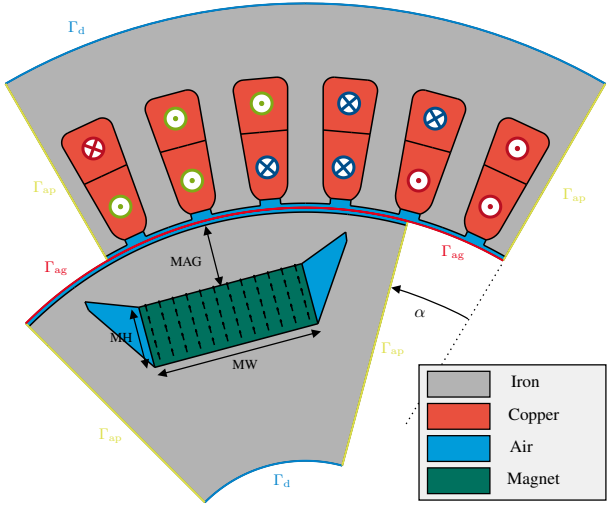}
    }
	\caption{Parametrization of the PMSM geometry including parameter names, material definitions and boundary conditions, \rwiii{based on} \cite{Komann_2024aa}.}
	\label{fig:PMSMgeometry}
\end{figure}

Let us consider the PMSM from \cite{Pahner_1998ab,Komann_2024aa} as depicted in Figure~\ref{fig:PMSMgeometry}. The rotor and stator iron cores are shown in gray, the (homogenized) copper slots in red, the rotor magnet in green and
the air gap and air pockets in blue. The  geometric design variables, e.g., $\mathbf{p} = [\operatorname{MH}, \operatorname{MW}, \operatorname{MAG}]$, will be varied later for surrogate modeling. In addition, the rotor’s rotation angle $\alpha$ is introduced explicitly and will be treated as a separate parameter in the learning task. For clarity, we define the full parameter vector as $\mathbf{P} = [\mathbf{p}, \alpha]$.

The field distribution in a laminated PMSM can be obtained by the magnetostatic approximation which neglects displacement and eddy currents \cite{Salon_1995aa, Rosu_2017aa}. The parameter-dependent magnetic vector potential $\mathbf{A}^{(\mathbf{P})}$ is then described by the PDE
\begin{equation}
    \label{eq:magnetostatic}
    \nabla \times (\nu \nabla \times \mathbf{A}^{(\mathbf{P})}) 
    = 
    \mathbf{J}_{\text{src}}^{(\mathbf{P})} 
    +
    \nabla \times (\nu \mathbf{B}_\text{rem}^{(\mathbf{P})})
\end{equation}
defined on the parametrized computational domain $\Omega^{(\mathbf{P})}$.
The magnetic flux density is given by $\mathbf{B}^{(\mathbf{P})} = \nabla \times \mathbf{A}^{(\mathbf{P})}$ and the possibly nonlinear reluctivity by $\nu = \nu(\cdot)$. The source current density  $\mathbf{J}_{\text{src}}^{(\mathbf{P})}$ and the remanent flux density $\mathbf{B}_\text{rem}^{(\mathbf{P})}$ of the permanent magnets may also depend on the parameter vector.
The model is simplified to a two-dimensional magnetostatic problem by presuming a sufficient axial length and neglecting three-dimensional effects, e.g., due to end windings. The governing equations are 
\begin{align}
        \label{eq:magnetostatics_2D_1}
        \nabla \cdot (\nu \nabla A_{z,\mathrm{R}}^{(\mathbf{P})}) &= \nu \nabla \cdot \mathbf{B}^{(\mathbf{P})}
        \quad&& \text{in } \Omega_{\mathrm{R}}^{(\mathbf{P})},
        \\
        \label{eq:magnetostatics_2D_2}
        \nabla \cdot (\nu \nabla A_{z,\mathrm{S}}^{(\mathbf{P})}) &= -J_{z,\text{src}}^{(\mathbf{P})} && \text{in } \Omega_{\mathrm{S}}^{(\mathbf{P})},
\end{align}
given the computational domains $\Omega_{\mathrm{R}}^{(\mathbf{P})}$ and $\Omega_{\mathrm{S}}^{(\mathbf{P})}$, for rotor and stator respectively. Note that (\ref{eq:magnetostatics_2D_1}--\ref {eq:magnetostatics_2D_2}) requires only the $z$-components of the magnetic vector potential and the source current density, i.e., $A_{z,\text{src}}^{(\mathbf{P})}$
and
$J_{z,\text{src}}^{(\mathbf{P})}$.
Problem (\ref{eq:magnetostatics_2D_1}--\ref {eq:magnetostatics_2D_2}) is complemented with homogeneous Dirichlet and anti-periodic boundary conditions, respectively applied to the boundaries denoted with $\Gamma_{\mathrm{d}}$ and $\Gamma_{\mathrm{ap}}$ in \autoref{fig:PMSMgeometry}. Continuity of the magnetic vector potential $A_z^{(\mathbf{P})}$ and the azimuthal magnetic field strength $H_\varphi^{(\mathbf{P})} =\nu_\mathrm{S} \nabla A_{z,\mathrm{S}}^{(\mathbf{P})}$ across $\Gamma_\text{ag}=\bar\Omega_{\mathrm{R}}^{(\mathbf{P})}\cap\bar\Omega_{\mathrm{S}}^{(\mathbf{P})}$ are enforced by a Lagrange multiplier and additional coupling conditions \cite{Egger_2022ab}.
Last, the source current density is given \mb{for each phase $k\in\left\{1,2,3\right\}$}
by a spatial uniform distribution \mb{with current $I_k$ and phase $\phi_0 +\frac{2 \pi}{3}k + \phi_k$}
within the coils  based on the stranded conductor model \cite{Schops_2013aa}.

In electric machines, one of the key performance indicators is the electromagnetic torque. The torque acting on a volume can be determined by integrating the Maxwell stress tensor over a surface that encloses this volume \cite{Salon_1995aa}. This can be simplified in the 2D context, such that the torque for a given parameter configuration $\mathbf{P}$ is given by
\begin{equation}
    T(A_z^{(\mathbf{P})}) = \frac{r^2L}{\mu_0}\int_0^{2\pi} B_r^{(\mathbf{P})} B_\varphi^{(\mathbf{P})} \mathrm{d} \varphi
    \label{eq:torque}
\end{equation}
with the radial and angular components of $\mathbf{B}^{(\mathbf{P})}$, i.e., $B_r^{(\mathbf{P})}$ and $B_\varphi^{(\mathbf{P})}$, the machine length $L$, the vacuum permeability $\mu_0$ and the radius $r$ of the integration line. The torque value depends on the parametric realization of the problem as the field does. The integration path is typically chosen within the air gap, making the air gap field sufficient to evaluate the torque.

\section{Methodology}

\label{sec:methodology}

The proposed methodology consists of three consecutive steps, i.e., space discretization by IGA, reduced basis construction by POD and the learning of modes by a DNN.

\subsection{IGA Discretization}
\begin{figure}
	\centering
	\scalebox{0.65}{\begin{tikzpicture}[y=0.80pt, x=0.80pt, yscale=-1, xscale=1, inner sep=0pt, outer sep=0pt,font=\Large]

    \path[draw=TUDa-9d,thick, fill=TUDa-9a, fill opacity=0.2] (20, 0) arc (-90:0:130) -- (85,130) arc (0:-90:65) -- cycle;

	\path[draw=black,thick, xshift=-3, yshift=3] (26.7600,114.8700) .. controls (40,95) and (47.3200,89.6700) .. (55,81)  node [midway, fill=white] {$\textbf{F}^{(\textbf{P}_1)}$};
	\path[fill=black, xshift=-3, yshift=3] (51, 76) -- (58, 87) -- (61, 78) -- cycle;
    \node[] at (135,30) {$\Omega^{(\textbf{P}_1)}$};

    \begin{scope}[shift={(-28,0)}]
    	\path[draw=TUDa-2d,thick, fill=TUDa-2a,fill opacity=0.2] (2,100+25) rectangle (2+50.0000,100+75);
    
    \foreach \i in {1,...,3}{
		\path[draw=TUDa-2d] (2+\i*50/4,125) -- (2+\i*50/4,175);
		\path[draw=TUDa-2d] (2,175-\i*50/4) -- (2+50,175-\i*50/4);
	}
  
    
    \foreach \i in {0,...,4}{
        \foreach \j in {0, ..., 4}{
            \node at (2+\i*50/4, 175 - \j*50/4) [circle, fill=black, inner sep=1pt] {};
        }
	}
      \end{scope}


     \foreach \i in {1,...,3}{
		\path[draw=TUDa-9d] (20, \i*65/4) arc (-90:0:130-\i*65/4);
		\path[draw=TUDa-9d] ({20+65*cos(-90+\i*90/4)},{130+65*sin(-90+\i*90/4)}) -- ({20+130*cos(-90+\i*90/4)},{130+130*sin(-90+\i*90/4)});
	}
    
    \foreach \i in {0,...,4}{
        \foreach \j in {0, ..., 4}{
            \node at ({20+(130-\i*65/4)*sin(\j*90/4)}, {130-(130-\i*65/4)*cos(\j*90/4)}) 
                [circle, fill=black, inner sep=1pt] {};
        }
	}
	

    \begin{scope}[rotate=-90,xslant=0.5, yslant=.4, scale=0.7, shift={(-120,-170)}]
    	
    	\path[draw=TUDa-9d,thick, fill=TUDa-9a, fill opacity=0.2] (20, 0) arc (-90:0:130) -- (85,130) arc (0:-90:65) -- cycle;
    	
    	 \foreach \i in {1,...,3}{
    		\path[draw=TUDa-9d] (20, \i*65/4) arc (-90:0:130-\i*65/4);
    		\path[draw=TUDa-9d] ({20+65*cos(-90+\i*90/4)},{130+65*sin(-90+\i*90/4)}) -- ({20+130*cos(-90+\i*90/4)},{130+130*sin(-90+\i*90/4)});
    	}
    	
    	   \foreach \i in {0,...,4}{
    		\foreach \j in {0, ..., 4}{
    			\node at ({20+(130-\i*65/4)*sin(\j*90/4)}, {130-(130-\i*65/4)*cos(\j*90/4)}) 
    			[circle, fill=black, inner sep=1pt] {};
    		}
    	}
    \end{scope}
    \begin{scope}[rotate=-90,shift={(-150,-150)}]
    \path[draw=black,thick, xshift=-3, yshift=3] (26.7600,114.8700) .. controls (40,95) and (47.3200,89.6700) .. (55,81)  node [midway, fill=white] {$\textbf{F}^{(\textbf{P}_2)}$};
    \path[fill=black, xshift=-3, yshift=3] (51, 76) -- (58, 87) -- (61, 78) -- cycle;
    \end{scope}
     \node[] at (-115,50) {$\Omega^{(\textbf{P}_2)}$};

%
%
	\node[] at (0,110) {$\hat\Omega$};
	
\end{tikzpicture}}
	\caption{Illustration of the parameter-dependent mapping from the reference domain $\hat{\Omega}$ to different physical domains $\Omega_1$ and~$\Omega_2$.}
	\label{fig:mapping}
\end{figure}
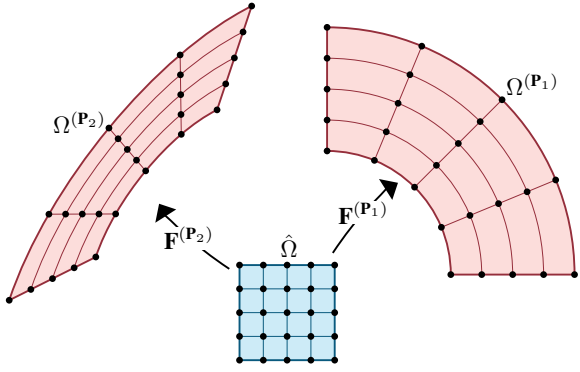

The PMSM's geometry representation is inspired by CAD. It is given by a map from the reference domain $\hat{\Omega} = \left[0,1\right]^d$ to a physical domain $\Omega^{(\mathbf{P})} \subset \mathbb{R}^r$, for our 2D model $d=r=2$. The standard tools for this representation are B-splines and NURBS.
Given a knot vector $\Xi \subset \left[0,1\right]$, 
the basis of univariate B-splines $\hat B_i^p$ of degree $p$ can be defined using the Cox-de-Boor recursion formula.%
From those one derives the NURBS basis functions $\hat{N}_i^p$ \cite{Cohen_2001aa}.
Curves are then described as linear combination of these functions, surfaces and higher-dimensional objects are created using tensor products~\cite{Cohen_2001aa}.
Using these constructions, it is possible to define maps~$\textbf{F}^{(\mathbf{P})}:~\hat{\Omega}~\rightarrow \Omega^{(\mathbf{P})}$ that described the parametrized physical domain such that $\mathbf{x} = \textbf{F}^{(\mathbf{P})}(\boldsymbol{\xi})$, $\mathbf{x} \in \Omega$, $\boldsymbol{\xi} \in \left[0,1\right]^d$. Note, for different geometrical realizations, the reference domain stays the same. That is essential if the parametric problem is learned in terms of its coefficients. For reasonable parametric changes, no remeshing is required and the learned coefficients belong to the same nodes in the reference domain which are then mapped to the corresponding physical domain, see \autoref{fig:mapping}.

Note that, in most practical applications, the geometry cannot be parametrized using a single map from the reference to the physical domain. This is also the case for the considered PMSM, which features a complex geometry with different material subdomains.
In such cases, a multi-patch parametrization is used, where the physical domain is decomposed into a collection of subdomains, each with a corresponding projection map, to be appropriately combined~\cite{Buffa_2015aa}.

Finally, IGA is used to discretize the problem (\ref{eq:magnetostatics_2D_1}--\ref {eq:magnetostatics_2D_2}) . Following the standard Ritz-Galerkin approach the B-splines are used to span the ansatz and test function spaces. The magnetic vector potential in the reference domain is approximated by
\begin{equation}
A_z^{(\mathbf{P})}(\textbf{x})
\approx 
\sum\nolimits_{i=1}^N B_i^p(\mathbf{P},\textbf{x})\; u_i
\label{eq:AnsatzTestFunctions}
\end{equation}
where  $u_i$ are the unknown coefficients and $B_i^p(\mathbf{P},\textbf{x})$ are the spline basis functions in the physical domain mapped by $\textbf{F}^{(\mathbf{P})}$.
Using harmonic mortaring for rotor-stator coupling \cite{Bontinck_2018ac} results in the matrix system
\begin{equation}
\underbrace{
\begin{pmatrix}
\mathbf{K}_\text{rt}(\mathbf{u_\text{rt}}) & \mathbf{0} & -\mathbf{G}_\text{rt} \\
\mathbf{0} & \mathbf{K}_\text{st} (\mathbf{u}_\text{st})& \mathbf{G}_\text{st} \mathbf{R}_{\alpha} \\
-\mathbf{G}_\text{rt}^\top & \mathbf{R}_{\alpha}^\top \mathbf{G}_\text{st}^\top & \mathbf{0} 
\end{pmatrix}
}_{\eqqcolon \mathbf{K}}
\underbrace{
\begin{pmatrix}
\mathbf{u}_\text{rt} \\
\mathbf{u}_\text{st} \\
\boldsymbol{\lambda} 
\end{pmatrix}
}_{\eqqcolon \mathbf{u}}
=
\underbrace{
\begin{pmatrix}
\mathbf{b}_\text{rt} \\
\mathbf{b}_\text{st} \\
\mathbf{0} 
\end{pmatrix}
}_{\eqqcolon \mathbf{b}},
\label{eq:system}
\end{equation}
where $\mathbf{K}_\text{rt}$, $\mathbf{K}_\text{st}$ are stiffness matrices and $\mathbf{G}_\text{rt}$, $\mathbf{G}_\text{st}$ coupling matrices for the rotor and stator, respectively, $\mathbf{R}_{\alpha} $ is the rotation matrix for a given rotation angle $\alpha$, $\boldsymbol{\lambda}$ is the vector of Lagrange multipliers, and $\mathbf{b}_\text{rt} $, $\mathbf{b}_\text{st} $ are the rotor- and stator-specific right-hand side vectors.  
For more details on constructing the matrix system \eqref{eq:system} the reader is referred to \cite{Wiesheu_2024aa}.
Note, that solution, matrices and right-hand sides all depend implicitly on the parameter vector $\mathbf{P}$, the system matrix also on the solution vector $\mathbf{u}$ due to the nonlinear \rwi{reluctivity of the iron material.} We denote the resulting torque from the finite element calculation by $\tau^{(\mathbf{P})}\approx T(A_z^{(\mathbf{P})})$.

\subsection{Proper Orthogonal Decomposition}
In this section, we briefly outline the proper orthogonal decomposition (POD) procedure used to extract a low-dimensional, physically interpretable basis from a set of high-fidelity simulation snapshots. First, the so-called snapshot matrix $\mathbf{S}$ is built by concatenating full-order solutions of the problem with different parameter configurations
\begin{equation}
\mathbf{S} =
\left[
\begin{array}{c c c c}
| & | &  & | \\
\mathbf{u}^{(\mathbf{P}_{1})} & \textbf{u}^{(\mathbf{P}_{2})} & \cdots & \mathbf{u}^{(\mathbf{P}_{M})}\\
| & | &  & | 
\end{array}
\right]
\in \mathbb{R}^{N\times M}
\end{equation}
with $\mathbf{u}^{(\mathbf{P})}=\bigl(\mathbf{u}_\text{rt}^{(\mathbf{P})},\mathbf{u}_\text{st}^{(\mathbf{P})}\bigr)$. The matrix has as many rows as DoFs ($N$) and as many columns as the number of experiments ($M$) performed.
The main assumption of the POD is that the finite element solution of the problem can be approximated by a linear combination of a small number $m\ll N$ of orthonormal vectors assembled in the matrix 
\begin{equation}
\mathbf{Q}_m =
\left[
\begin{array}{c c c c}
| & | &  & | \\
\mathbf{q}_1 & \textbf{q}_2 & \cdots & \textbf{q}_3\\
| & | &  & | 
\end{array}
\right]
\in \mathbb{R}^{N\times m}.
\end{equation}
The reduced basis $\mathbf{Q}_m$ is commonly obtained by performing a truncated singular value decomposition (SVD) of $\mathbf{S = U\Sigma V}$, using only the first $m$ vectors of $\mathbf{U}$~\cite{Benner_2017aa}. However, instead of performing a SVD, 
one can compute the eigenvectors $\bm{\phi}_1, \ldots, \bm{\phi}_m \in \mathbb{R}^N$ corresponding to the $m$ eigenvalues of largest magnitude from a symmetric $M \times M$ eigenvalue problem. In the finite element context, 
it is recommended to perform the POD with an appropriately weighted inner product~\cite{Volkwein_2013aa}. Let us endow the Euclidean space with the weighted inner product
\begin{equation}
    \langle \bm{\phi}, {\bm{\phi}'} \rangle_\mathbf{W} = \bm{\phi}^{\!\top} \mathbf{W} {\bm{\phi}'}
    \label{eq:eig-weighted}
\end{equation}
where $\mathbf{W} \in \mathbb{R}^{n\times n}$ is a symmetric, positive definite matrix. Then we can compute
\begin{equation}
    \mathbf{S}^{\!\top} \mathbf{W} \mathbf{S} \, \bar{\bm{\phi}}_i = \lambda_i \bar{\bm{\phi}}_i \quad \text{for} \ i=1,\ldots,k
    \label{eq:modes-weighted}
\end{equation}
and set
\begin{equation}
    \mathbf{q}_i = \frac{1}{\sqrt{\lambda_i}} \mathbf{S} \bar{\bm{\phi}}_i \quad \text{for} \ i=1,\ldots,k.
\end{equation}
In the magnetostatic setting, the solution space is the Sobolev space $\mathbf{H}(\mathrm{curl}; \Omega^{(\mathbf{P})})$ \cite{Monk_2003aa}. The corresponding semi-norm reads 
\begin{equation}
    \langle \mathbf{u}, \mathbf{v} \rangle_{\mathbf{H}(\mathrm{curl})} 
= \int_{\Omega^{(\mathbf{P})}} (\nabla \times \mathbf{u}) \cdot (\nabla \times \mathbf{v}) \, \mathrm{d}\mathbf{x}.
\end{equation}
Correspondingly, the discrete system inherits this inner product through the stiffness matrix \rwi{$\mathbf{K}$}, constructed with homogeneous unit material, associated with the curl-curl operator. 
This makes it a natural choice for defining the weighting matrix in the POD \rwi{since it establishes a relation to magnetic energy and compensates, for example, mesh inhomogeneities.} Note that the stiffness matrix depends on the parametrization $\mathbf{P}$, but we use a fixed weighting matrix \rwi{$\bar{\mathbf{K}}$} corresponding to the expected value $\bar{\mathbf{P}}$ and rotation angle $\alpha = 10$ degrees. Finally, we project onto the reduced basis and reconstruct the coefficients by
\begin{equation}
    \bar{\mathbf{u}}^{(\mathbf{P})} = \mathbf{Q}_m^{\top}  \rwi{\bar{\mathbf{K}}}
    \mathbf{u}^{(\mathbf{P})}
    \quad
    \text{and}
    \quad
    \label{eq:reconstruct}
    \mathbf{u}^{(\mathbf{P})}\approx \mathbf{Q}_m \bar{\mathbf{u}}^{(\mathbf{P})},
\end{equation}
respectively. Using this approach each solution can be stored using $m \ll N$ coefficients, together with the basis $\mathbf{Q}_m$. Selecting the number $m$ related to the eigenvalues with largest magnitude involves a trade-off between model complexity and accuracy: the more modes included, the better the reconstruction accuracy, but also the larger the computational effort.
A common approach \cite{Volkwein_2013aa} is to choose $m$ so that the so-called relative
cumulative energy is greater than a certain tolerance $\textrm{tol}$, i.e.,
\begin{equation}
    \epsilon_{\lambda} = 
    \frac{\sum_{i=1}^m {\lambda_i}}{\sum_{i=1}^{M} \lambda_i} \geq \textrm{tol}.
\end{equation}
We also define the relative reconstruction error in terms of the same semi-norm, namely
\begin{equation}
    \epsilon_\mathrm{rel, POD}^2 = 
    \frac{ \left(\mathbf{u} - \mathbf{Q}_m\bar{\mathbf{u}} \right)^{\!\top} \rwi{\mathbf{\bar{K}}} \left(\mathbf{u} - \mathbf{Q}_m \bar{\mathbf{u}} \right)}{\mathbf{u}^{\!\top}
    \rwi{\mathbf{\bar{K}}} \bar{\mathbf{u}}}.
    \label{eq:POD_err}
\end{equation}
In practice, we choose $m$ such that the POD error remains below the prediction error of the DNN. This ensures that the overall surrogate accuracy is governed by the learning model rather than the basis reduction.

\subsection{Neural Networks}
\begin{figure}
     \centering
     \begin{tikzpicture}[scale=.5]
  \node[anchor=east] at (0,0) {$\mathbf{P}$};
  
  \draw[->, ultra thick] (0,0) -- (2,0) node[midway, above] {};
  
  \draw (2, -1) rectangle (5, 1) node[pos=.5] {DNN};
  
  \draw[->, ultra thick] (5, 0) -- (7,0) node[midway, above] {$\bar{\mathbf{u}}^{(\mathbf{P})}$};
  
  \draw (7, -1) rectangle (11, 1) node[pos=.5] {POD \eqref{eq:reconstruct}};

  \draw[->, ultra thick] (11, 0) -- (13,0) node[midway, above] {$\textbf{u}^{(\mathbf{P})}$};

  \node[anchor=west] at (13,0) {$A_z^{(\mathbf{P})}(\textbf{x})$};
  
\end{tikzpicture}

     \caption{This flowchart illustrates the steps that the POD-DNN surrogate model makes to predict the coefficients for an new parametrization $\mathbf{P}.$}
     \label{fig:POD-DNN}
 \end{figure}
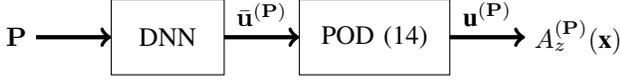

In this section we present the physics-informed machine learning method for learning 2-D parametric magnetostatic problems. The goal is to train a DNN that predicts the magnetic field solution of the PMSM as a function of its design parameters $\mathbf{P}$. 
Instead of predicting field values at discrete points, the network outputs the coefficients $\mathbf{u}$ of the magnetic vector potential in the isogeometric basis \eqref{eq:AnsatzTestFunctions}. This means that the learned solution can be reconstructed as a continuous function, fully embedded in the space spanned by the B-spline basis functions, maintaining physical plausibility such as the correct continuity conditions across interfaces. To keep the representation efficient, the DNN predicts a reduced number of coefficients in a POD basis, which are then projected back onto the full space (see \autoref{fig:POD-DNN}).

The network architecture consists of fully-connected layers. The parametric information is passed through the net by multiplying with the weights $\theta$ and applying activation functions~$\sigma$~\cite{Goodfellow_2016aa}. The input layer takes the parameter vector $\mathbf{P}$; the output layer provides the reduced coefficient vector $\bar{\mathbf{u}}$. Hyperparameters such as the number of layers and neurons, learning rate, regularization strength, and the number of training epochs are tuned for optimal performance.

Training is physics-informed by incorporating the underlying parametric magnetostatic problem into the loss function. Specifically, the loss measures the semi-norm-based relative error of the magnetic field $\mathbf{B}$
\begin{equation}
 \label{eq:loss}
   \epsilon_\mathrm{rel, DNN}^2
   =
   \frac{
       \left(\bar{\mathbf{u}} - \tilde{\mathbf{u}}\right)^{\top} 
       \bar{\mathbf{K}}
       \left(\bar{\mathbf{u}} - \tilde{\mathbf{u}}\right)
   }{ 
       \bar{\mathbf{u}}^{\top} 
       \bar{\mathbf{K}}
       \bar{\mathbf{u}}
    }.
\end{equation} 
where $\bar{\mathbf{u}}$ are the IGA coefficients of the training sample projected onto the POD basis, $\bar{\mathbf{K}}$ is again the fixed weighting matrix and $\tilde{\mathbf{u}}$ are the network's prediction. Although this approximates the true error in the IGA space, it significantly reduces computational cost and remains accurate, provided the POD space is sufficiently rich and the stiffness matrix varies moderately across the parameter space. \rwiii{Training is  terminated once the loss function values fall below some tolerance}. Regularization and normalization of the data are applied to prevent overfitting and ensure stable training. After training, the accuracy of the surrogate is assessed on a testing dataset by computing the error in the full IGA space, using the dedicated stiffness matrix for each parametric realization.

\section{Numerical Results}

\label{sec:results}
In this section, we present numerical results for two surrogate models: 
one learning the magnetic field in the air gap only, and one predicting the full magnetic field in rotor, stator, and air gap regions. For each case, we detail the POD basis selection, neural network architecture and training procedure, accuracy assessment, and runtime performance. The design space of the parametric problem is given in \autoref{tab:parameter_ranges}. 

The DNN's weights $\theta$ are adapted using the optimizer ADAM \cite[Sec.~8.3, 8.5]{Goodfellow_2016aa} to minimize the sum of $\varepsilon_{\textrm{rel}}$ over all training samples. Hyperparameter optimization was performed using Optuna on the validation samples~\cite{Akiba_2019aa}.
All training, testing, and validation samples are generated with a high-fidelity isogeometric FEM model of the PMSM. The parameter space is sampled using Quasi-Monte Carlo sampling yielding $1024$~training samples and $128$ each for testing and validation.
For implementation of the DNN the Python library PyTorch~\cite{Paszke_2019aa} was used.
All computations were run on a consumer laptop from Intel\textsuperscript{\textregistered} Core\textsuperscript{TM} Ultra 7 155H CPU (16 cores / 22 threads), 32GB RAM, Intel Arc Graphics (Meteor Lake-P).

\begin{figure}
    \centering
    \input{./images/airgap_reconstruction_error.tex}
    \vspace{-1em}
    \caption{Decay of $\epsilon_\mathrm{rel, POD} $ for air gap and full field snapshots. Reconstruction errors of $0.1\%$ and $0.6\%$ are marked with dashed grey line, selected number of modes in red. }
    \label{fig:rce_airgap}
\end{figure}

\begin{table}
\centering
\caption{Parameter ranges for field learning.}
\begin{tabular*}{0.8\linewidth}{@{\extracolsep{\fill}}lcc}
\hline
\textbf{Parameter} & \textbf{Min} & \textbf{Max} \\
\hline
MH      & \SI{1.5}{\milli \meter} & \SI{12}{\milli\meter} \\
MW      & \SI{7}{\milli \meter}   & \SI{23}{\milli \meter} \\
MAG     & \SI{5}{\milli \meter}  & \SI{15}{\milli\meter} \\
$\alpha$  & $0^{\circ}$                  & $20^{\circ}$ \\
LW1 &  \SI{1}{\milli \meter}   & \SI{3}{\milli \meter} \\
LW2 & \SI{2}{\milli \meter}   & \SI{4}{\milli \meter} \\
Theta1 & $18^{\circ}$                  & $22^{\circ}$ \\
Theta2  & $0.1^{\circ}$                  & $0.6^{\circ}$ \\
$I_0$, $I_1$, $I_2$ & \SI{9}{\ampere} & \SI{11}{\ampere} \\
$\phi_0$, $\phi_1$,$\phi_2$  & $-5^{\circ}$ &$5^{\circ}$\\
\hline
\end{tabular*}
\label{tab:parameter_ranges}
\end{table}

\subsection{Surrogate for Air gap Field}

To extract a low-dimensional representation of the air gap field, we first perform POD on the snapshot matrix comprising $1024$ high-fidelity solutions,~i.e. the training samples and with $4$ input parameteres, i.e. $\mathbf{P} = [\operatorname{MH}, \operatorname{MW}, \operatorname{MAG},  \alpha]$. \autoref{fig:rce_airgap} shows the decay of the mean relative reconstruction error $\epsilon_\mathrm{rel, POD}$ of the validation samples. 
We select $k = 14$ modes for $n = 462$ DoFs, which results in a mean relative POD reconstruction error of~$\epsilon_\mathrm{rel, POD} < 0.1\,\%$.

The reduced-order coefficients are mapped to the parameter space 
using a feedforward neural network with the architecture summarized 
in \autoref{tab:nn_airgap}. 
\begin{table}
    \centering
    \caption{Neural network architecture for air gap surrogate.}
    \label{tab:nn_airgap}
    \begin{tabular*}{0.8\linewidth}{@{\extracolsep{\fill}} l c}
    \hline
    \textbf{Parameter} & \textbf{Value} \\ \hline
    Number of layers & $3$ \\
    Neurons per layer & $[190,\,110,\,180]$ \\
    Activation function & ReLU \\
    \rwiii{Tolerance} & $3\times10^{-5}$ \\
    Training epochs & 10{,}234 \\
    \mb{Training time} & \mb{\SI{45}{\second}}. \\
    \hline
  \end{tabular*}
\end{table}

\autoref{tab:errors_airgap} shows the mean relative errors on the training, test and validation sets measured in the full IGA space. The POD truncation error is below the approximation error of the neural network, ensuring that the dimensionality reduction does not limit accuracy. \rwii{Training without POD results in similar errors, however it takes five times longer.}

\begin{table}
    \centering
    \caption{Relative errors (\%) for the air gap surrogate.}
    \label{tab:errors_airgap}
    \begin{tabular*}{0.8\linewidth}{@{\extracolsep{\fill}}lccc}
        \hline
        \textbf{Dataset} & \textbf{Mean} & \textbf{Max} & \textbf{Std. dev.} \\ \hline
        Training set & 0.55 & 1.35 & 0.11 \\
        Test set & 0.62 & 2.16 & 0.24 \\
        Validation set & 0.60 & 1.54 & 0.17 \\
        \hline
    \end{tabular*}
\end{table}

As an application-oriented metric, the mean relative error of the torque across the design space specified in \autoref{tab:parameter_ranges} is $1.3\,\%$ in comparison to the full-order model.%
The average runtime for a full IGA solve is approximately \SI{9}{\second}, while the surrogate prediction (including POD projection and DNN inference) takes \SI{0.3}{\milli\second}. This results in a speed-up factor of approximately~$30000$. 

\subsection{Surrogate for Full Field}
\begin{figure*}
    \centering
    \begin{subfigure}[b]{0.5\textwidth}
        \centering
        \vspace{-0.35cm}
        \includegraphics
        [width=0.65\textwidth]{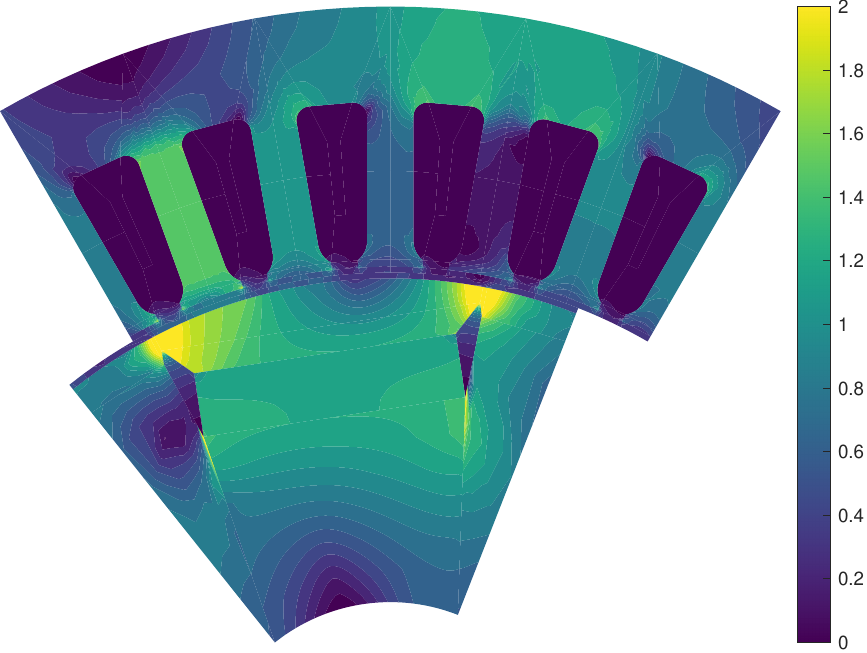}
        \vspace{-0.1cm}
        \caption{Predicted magnetic flux density $\mathbf{B}^{(\mathbf{P})} \,  [\si{\tesla}]$. }
        \label{fig:field_full_a}
    \end{subfigure}%
    \begin{subfigure}[b]{0.5\textwidth}
        \centering
        \vspace{-0.35cm}
        \includegraphics
        [width=0.65\textwidth]{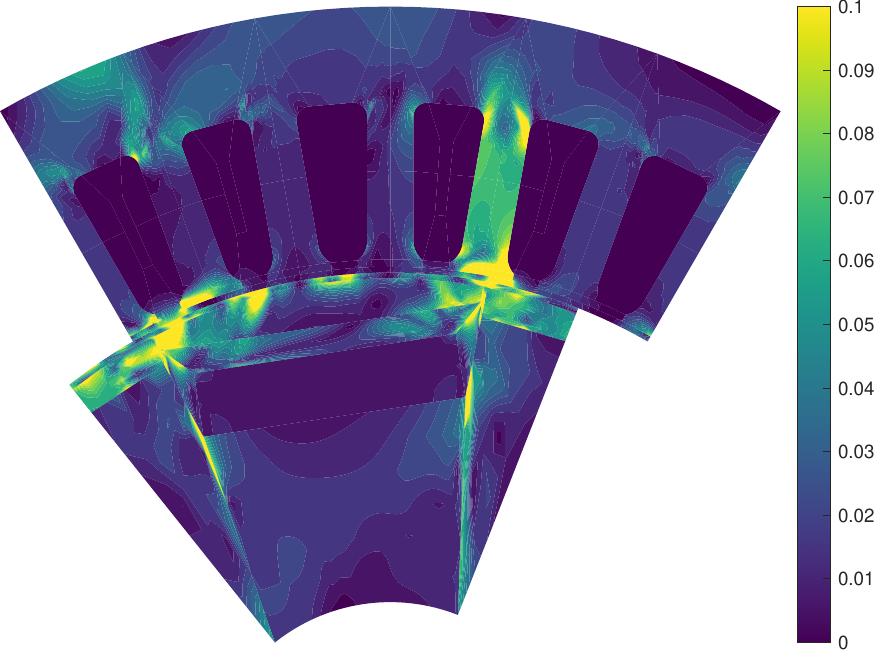}
        \vspace{-0.1cm}
        \caption{Absolute error in the magnetic flux density  [\si{\tesla}] compared to high-fidelity solution.}
        \label{fig:field_full_b}
    \end{subfigure}

    \caption{Reconstructed magnetic field and absolute error for~$\mathbf{P}=$[$\SI{5.55}{\milli\meter}$, $\SI{22.90}{\milli\meter}$, $\SI{6.08}{\milli\meter}$, $\ang{8.58}$, $\SI{1.39}{\milli\meter}$, $\SI{2.57}{\milli\meter}$, $\ang{19.09}$, $\ang{0.49}$, $\SI{10.29}{\ampere}$, $\SI{9.01}{\ampere}$, $\SI{9.6}{\ampere}$, $\ang{-0.9}$, $\ang{2.95}$, $\ang{-2.67}$] with relative error of~$2.83\%$.}
    \label{fig:field_full}
\end{figure*}

For the full geometry, the snapshot matrix includes the rotor, stator, and air gap fields of the training samples. \mb{Moreover, the input dimension is increased from $4$ to $14$ parameters, see \autoref{tab:parameter_ranges}. We choose $k = 90$ modes out of $n = 6354$ DoFs, resulting in a mean relative POD reconstruction error of the validation samples of $\epsilon_\mathrm{rel} < 0.6\,\%$ (see \autoref{fig:rce_airgap}).} An increased reconstruction error is acceptable, since it aligns with the anticipated higher prediction errors of the network.
The network has the architecture summarized 
in \autoref{tab:nn_full}.
\begin{table}
    \centering
    \caption{Network architecture for full field surrogate.}
    \label{tab:nn_full}
    \begin{tabular*}{0.8\linewidth}{@{\extracolsep{\fill}} l c}
        \hline
        \textbf{Parameter} & \textbf{Value} \\ \hline
        Number of layers &  3 \\
        Neurons per layer & 336 \\
        Activation function & ReLU \\
        \rwiii{Tolerance} & $2.2\times10^{-4}$\\
        \mb{Training epochs} & \mb{$8{,}100$} \\
        \mb{Training time} &  \mb{\SI{165}{\second}.} \\
        \hline
    \end{tabular*}
\end{table}
\autoref{tab:errors_full} lists the errors on the training, validation, and test sets. \autoref{fig:field_full} shows a representative field prediction and its absolute error. The mean error of the torque is \mb{$2.5\,\%$} in comparison to the full-order model. \mb{Considering 4 input parameters as in the air gap field case, results in a mean relative error of $1.5\%$. In the case of 14 parameters, the error rises slightly, indicating that the model generalizes well despite the added complexity.}
\begin{table}
    \centering
    \caption{Relative errors (\%) for the full
    field surrogate.}
    \label{tab:errors_full}
    \begin{tabular*}{0.8\linewidth}{@{\extracolsep{\fill}} lccc}
     \hline
        \textbf{Dataset} & \textbf{Mean} & \textbf{Max} & \textbf{Std. dev.} \\ \hline
        Training set & 1.78 & 9.56 & 0.92 \\
        Validation set & 2.77 & 6.57 & 1.19 \\
        Test set & 2.84 & 9.43 & 1.38 \\
        \hline
    \end{tabular*}
\end{table}

The average runtime for a full IGA solve is again approximately \SI{9}{\second}. The full
field surrogate requires \SI{50}{\milli \second} per prediction, resulting in a speed-up factor of about $180$. Again, the POD truncation error is smaller than the DNN prediction error, ensuring efficient dimensionality reduction. \rwii{Training without POD took about $\SI{4}{\hour}$ with $1.3 \times$ higher errors, demonstrating the effectiveness of the reduced basis.}

\subsection{Discussion}

In summary, \rwiii{the construction of both surrogate models pays off after approximately 1,300 evaluations: 1,280 samples were used for training, testing, and validation, and the training time equivalents about 5 evaluations for the airgap field and 16 for the full field. As the network evaluation takes only milliseconds, its computational cost is almost negligible.} From this point onward, the air gap surrogate achieves a higher runtime gain 
due to a smaller reduced basis and neural network, and is suitable 
for applications focusing on derived quantities like torque. 
The full
field surrogate allows to reconstruct all DoFs, 
making it preferable when field information in rotor or stator 
is also required. The choice depends on the balance between computational 
cost and required physical detail. \mb{Despite increasing the input space, the prediction errors remain low. POD helps in decreasing training times and errors.}

\section{Conclusion}

\label{sec:conclusion}
In this work, we presented a hybrid reduced-order modeling approach combining POD with DNNs to efficiently approximate the magnetostatic field in a parametric model of a rotating electrical machine. The methodology exploits the physical structure of the problem by constructing a low-dimensional basis using POD, while the DNN learns the non-linear parametric dependence of the solution coefficients in the reduced space.

Two representative cases were investigated: learning only the air gap field and learning the full magnetic field distribution across rotor, stator, and air gap. For both cases, we demonstrated that the number of POD modes can be chosen such that the approximation error introduced by the projection remains below the prediction error of the DNN, ensuring no significant loss of accuracy while substantially reducing training cost.

We systematically evaluated the performance of the trained networks in terms of relative errors on training, validation, and test data, showing good generalization across the parameter space. Additionally, the impact on a derived engineering quantity, the electromagnetic torque, was analyzed, confirming that the hybrid model retains sufficient accuracy for practical use. A notable runtime gain was achieved compared to the high-fidelity finite element solution, highlighting the potential for rapid design iterations and real-time applications.

Overall, the results indicate that the proposed POD–DNN framework is a promising approach for parametric field prediction in complex geometries where conventional high-fidelity models are too computationally demanding.

\section*{Acknowledgment}
\footnotesize

The author are grateful for the support of Felicia Bossani and thank Stefan Kurz, Matthias Möller and Brahim Ramdane for many fruitful discussions. Support of the CRC CREATOR (TRR 361), the Graduate School CE at TU Darmstadt and the IRGA program of Université Grenoble Alpes is acknowledged.



\begin{thebibliography}{10}
\providecommand{\url}[1]{#1}
\csname url@samestyle\endcsname
\providecommand{\newblock}{\relax}
\providecommand{\bibinfo}[2]{#2}
\providecommand{\BIBentrySTDinterwordspacing}{\spaceskip=0pt\relax}
\providecommand{\BIBentryALTinterwordstretchfactor}{4}
\providecommand{\BIBentryALTinterwordspacing}{\spaceskip=\fontdimen2\font plus
\BIBentryALTinterwordstretchfactor\fontdimen3\font minus
  \fontdimen4\font\relax}
\providecommand{\BIBforeignlanguage}[2]{{%
\expandafter\ifx\csname l@#1\endcsname\relax
\typeout{** WARNING: IEEEtran.bst: No hyphenation pattern has been}%
\typeout{** loaded for the language `#1'. Using the pattern for}%
\typeout{** the default language instead.}%
\else
\language=\csname l@#1\endcsname
\fi
#2}}
\providecommand{\BIBdecl}{\relax}
\BIBdecl

\bibitem{Alizadeh_2020aa}
R.~Alizadeh, J.~K. Allen, and F.~Mistree, ``Managing computational complexity
  using surrogate models: a critical review,'' \emph{Research in Engineering
  Design}, vol.~31, no.~3, pp. 275--298, 2020.

\bibitem{Parekh_2023aa}
V.~Parekh, D.~Flore, and S.~Schöps, ``\BIBforeignlanguage{english}{Performance
  analysis of electrical machines using a hybrid data- and physics-driven
  model},'' \emph{\BIBforeignlanguage{english}{{IEEE} Trans. Energ. Convers.}},
  vol.~38, no.~1, 2023.

\bibitem{Salon_1995aa}
S.~J. Salon, \emph{\BIBforeignlanguage{english}{Finite Element Analysis of
  Electrical Machines}}.\hskip 1em plus 0.5em minus 0.4em\relax Kluwer, 1995.

\bibitem{Rosu_2017aa}
M.~Rosu, P.~Zhou, D.~Lin, D.~M. Ionel, M.~Popescu, F.~Blaabjerg, V.~Rallabandi,
  and D.~Staton, \emph{\BIBforeignlanguage{english}{Multiphysics Simulation by
  Design for Electrical Machines, Power Electronics and Drives}}.\hskip 1em
  plus 0.5em minus 0.4em\relax Wiley Press, 2017.

\bibitem{Raissi_2019aa}
M.~Raissi, P.~Perdikaris, and G.~E. Karniadakis, ``Physics-informed neural
  networks: A deep learning framework for solving forward and inverse problems
  involving nonlinear partial differential equations,'' \emph{J. Comput.
  Phys.}, vol. 378, pp. 686--707, 2019.

\bibitem{Beltran-Pulido_2022aa}
A.~Beltrán-Pulido, I.~Bilionis, and D.~Aliprantis, ``Physics-informed neural
  networks for solving parametric magnetostatic problems,'' \emph{{IEEE} Trans.
  Energ. Convers.}, vol.~37, no.~4, pp. 2678--2689, 2022.

\bibitem{von-Tresckow_2022aa}
M.~von Tresckow, S.~Kurz, H.~De~Gersem, and D.~Loukrezis, ``A neural solver for
  variational problems on {CAD} geometries with application to electric machine
  simulation,'' \emph{J. Mach. Learn. Mod. Comput.}, vol.~3, no.~1, pp. 49--75,
  2022.

\bibitem{Monk_2003aa}
P.~Monk, \emph{\BIBforeignlanguage{english}{Finite Element Methods for
  {Maxwell}'s Equations}}.\hskip 1em plus 0.5em minus 0.4em\relax Oxford:
  Oxford University Press, 2003.

\bibitem{Zorzetto_2025aa}
M.~Zorzetto, R.~Torchio, F.~Lucchini, M.~Forzan, and F.~Dughiero, ``Proper
  orthogonal decomposition for parameterized macromodeling of a longitudinal
  electromagnetic levitator,'' \emph{{IEEE} Trans. Magn.}, vol.~61, no.~4, pp.
  1--7, 2025.

\bibitem{Moller_2021aa}
M.~Möller, D.~Toshniwal, and F.~van Ruiten, ``Physics-informed machine
  learning embedded into isogeometric analysis,'' in \emph{Mathematics: Key
  Enabling Technology for Scientific Machine Learning},
  2021, pp. 57--59.

\bibitem{Backmeyer_2024ab}
M.~Backmeyer, S.~Kurz, M.~Möller, and S.~Schöps, ``Solving electromagnetic
  scattering problems by isogeometric analysis with deep operator learning,''
  in \emph{2024 Kleinheubach Conference}.\hskip 1em plus 0.5em minus
  0.4em\relax IEEE, 2024.

\bibitem{Evans_2009aa}
J.~A. Evans, Y.~Bazilevs, I.~Babuška, and T.~J.~R. Hughes, ``n-widths,
  sup–infs, and optimality ratios for the k-version of the isogeometric
  finite element method,'' \emph{Comput. Meth. Appl. Mech. Eng.}, vol. 198, no.
  21-26, pp. 1726--1741, 2009.

\bibitem{Cohen_2001aa}
E.~Cohen, R.~F. Riesenfeld, and G.~Elber, \emph{Geometric Modeling with
  Splines: An Introduction}.\hskip 1em plus 0.5em minus 0.4em\relax CRC Press,
  2001.

\bibitem{Buffa_2015aa}
A.~Buffa, R.~Vázquez~Hernández, G.~Sangalli, and L.~Beirão~da Veiga,
  ``\BIBforeignlanguage{english}{Approximation estimates for isogeometric
  spaces in multipatch geometries},'' \emph{\BIBforeignlanguage{english}{Numer.
  Meth. Part. Differ. Equat.}}, vol.~31, no.~2, pp. 422--438, 2015.

\bibitem{Benner_2017aa}
P.~Benner, M.~Ohlberger, A.~Patera, G.~Rozza, and K.~Urban, Eds.,
  \emph{\BIBforeignlanguage{english}{Model Reduction of Parametrized
  Systems}}.\hskip 1em plus 0.5em minus 0.4em\relax Springer, 2017.

\bibitem{Swischuk_2019aa}
R.~Swischuk, L.~Mainini, B.~Peherstorfer, and K.~Willcox, ``Projection-based
  model reduction: Formulations for physics-based machine learning,''
  \emph{Computers \& Fluids}, vol. 179, pp. 704--717, 2019.

\bibitem{Henneron_2020aa}
T.~Henneron, A.~Pierquin, and S.~Clénet, ``Surrogate model based on the {POD}
  combined with the {RBF} interpolation of nonlinear magnetostatic {FE}
  model,'' \emph{{IEEE} Trans. Magn.}, vol.~56, no.~1, pp. 1--4, 2020.

\bibitem{Fresca_2022aa}
S.~Fresca and A.~Manzoni, ``{POD}-{DL}-{ROM}: Enhancing deep learning-based
  reduced order models for nonlinear parametrized {PDEs} by proper orthogonal
  decomposition,'' \emph{Comput. Meth. Appl. Mech. Eng.}, vol. 388, p. 114181,
  2022.

\bibitem{Zorzetto_2025ab}
\BIBentryALTinterwordspacing
M.~Zorzetto, R.~Torchio, F.~Lucchini, P.~D. Barba, M.~E. Mognaschi, M.~Forzan,
  and F.~Dughiero, ``\BIBforeignlanguage{english}{Machine learning-based
  reduced order modeling of nonlinear magnetic devices},'' 2025, submitted to IEEE
  Transactions on Magnetics.
\BIBentrySTDinterwordspacing

\bibitem{Volkwein_2013aa}
S.~Volkwein, \emph{\BIBforeignlanguage{english}{Proper Orthogonal
  Decomposition: Theory and Reduced-Order Modelling}}.\hskip 1em plus 0.5em
  minus 0.4em\relax University of Konstanz, 2013.

\bibitem{Pahner_1998ab}
U.~Pahner, ``\BIBforeignlanguage{english}{A general design tool for the
  numerical optimisation of electromagnetic energy transducers},'' PhD Thesis,
  KU Leuven, 1998.

\bibitem{Komann_2024aa}
T.~Komann, M.~Wiesheu, S.~Ulbrich, and S.~Schöps,
  ``\BIBforeignlanguage{english}{Robust design optimization of electric
  machines with isogeometric analysis},''
  \emph{\BIBforeignlanguage{english}{Math.}}, vol.~12, no.~9, 2024.

\bibitem{Egger_2022ab}
H.~Egger, M.~Harutyunyan, R.~Löscher, M.~Merkel, and S.~Schöps, ``On torque
  computation in electric machine simulation by harmonic mortar methods,''
  \emph{J. Math. Ind.}, vol.~12, no.~6, 2022.

\bibitem{Schops_2013aa}
S.~Schöps, H.~De~Gersem, and T.~Weiland,
  ``\BIBforeignlanguage{english}{Winding functions in transient
  magnetoquasistatic field-circuit coupled simulations},''
  \emph{\BIBforeignlanguage{english}{{COMPEL}}}, vol.~32, no.~6, pp.
  2063--2083, 2013.

\bibitem{Bontinck_2018ac}
Z.~Bontinck, J.~Corno, S.~Schöps, and H.~De~Gersem,
  ``\BIBforeignlanguage{english}{Isogeometric analysis and harmonic
  stator-rotor coupling for simulating electric machines},''
  \emph{\BIBforeignlanguage{english}{Comput. Meth. Appl. Mech. Eng.}}, vol.
  334, pp. 40--55, 2018.

\bibitem{Wiesheu_2024aa}
M.~Wiesheu, T.~Komann, M.~Merkel, S.~Schöps, S.~Ulbrich, and I.~Cortes~Garcia,
  ``Combined parameter and shape optimization of electric machines with
  isogeometric analysis,'' \emph{Optim. Eng.}, 2024.

\bibitem{Goodfellow_2016aa}
I.~Goodfellow, Y.~Bengio, and A.~Courville, \emph{Deep learning}.\hskip 1em
  plus 0.5em minus 0.4em\relax MIT press, 2016.

\bibitem{Akiba_2019aa}
T.~Akiba, S.~Sano, T.~Yanase, T.~Ohta, and M.~Koyama, ``{Optuna: A
  Next-generation Hyperparameter Optimization Framework},'' in
  \emph{Proceedings of the 25th ACM SIGKDD International Conference on
  Knowledge Discovery {\&} Data Mining (KDD '19)}.\hskip 1em plus 0.5em minus
  0.4em\relax ACM, 2019, pp. 2623--2631.

\bibitem{Paszke_2019aa}
A.~Paszke et al., ``Pytorch: An imperative style, high-performance deep learning
  library,'' in \emph{Advances in Neural Information Processing Systems 32
  ({NeurIPS} 2019)}.\hskip 1em plus 0.5em minus 0.4em\relax Curran Associates,
  Inc., 2019.

\end{thebibliography}
\end{document}